\documentclass[a4paper,UKenglish]{lipics}
\usepackage{amsmath}
\usepackage{amsfonts}
\usepackage{amssymb}
\usepackage{verbatim}
\usepackage{algorithmic}
\usepackage{algorithm}

\usepackage{lscape}
\usepackage{tabularx}
\usepackage{url}

\newcolumntype{I}{!{\vrule width 3pt}}
\newlength\savedwidth

\def\dd{\mathinner{.\,.}}
\newcommand{\cO}{\mathcal{O}}

\newtheorem{problem}{Problem}

\begin{document}

\title{On the Average-case Complexity of Pattern Matching with Wildcards}

\author{Carl Barton}
\affil{Blizard Institute, Bart's and the London School of Medicine and Dentistry, Queen Mary University of London, UK\\
{c.barton@qmul.ac.uk} 
}

\maketitle


\begin{abstract}
Pattern matching with wildcards is the problem of finding all factors of a text $t$ of length $n$ that match a pattern $x$ of length $m$, where wildcards (characters that match everything) may be present. In this paper we present a number of fast average-case algorithms for pattern matching where wildcards are restricted to either the pattern or the text, however, the results are easily adapted to the case where wildcards are allowed in both. We analyse the \textit{average-case} complexity of these algorithms and show the first non-trivial time bounds. These are the first results on the average-case complexity of pattern matching with wildcards which, as a by product, provide with first provable separation in time complexity between exact pattern matching and pattern matching with wildcards in the word RAM model.
\end{abstract}

\section{Introduction}
Pattern matching with wildcards is a string matching problem where the alphabet consists of standard letters and a \textit{wildcard} $\phi$ which matches every character in the alphabet. Given a text $t$ of length $n$ and a pattern $x$ of length $m<n$ the problem consists of finding all factors of the text that match the pattern.
Pattern matching with wildcards naturally arises in a number of problems in bioinformatics, they are primarily used to model \textit{single nucleotide polymorphisms} (SNP), being used in the identification of diseases and in genome wide association studies as markers for gene mapping. 

An early result in pattern matching with wildcards was the \textit{Fast Fourier Transform} (FFT) based algorithm of Fischer and Paterson~\cite{Fischer:1974:SOP:889566} with runtime $\cO(n \log m \log \sigma)$ for an alphabet of size $\sigma$. A subsequent study by Pinter~\cite{Pinter} outlined why the intransitivity of the match relation prevents algorithms such as KMP~\cite{Knuthf74fastpattern} being used or easily modified when wildcards are present. After Fischer and Patterson, much work focused on improving the algorithms by removing the dependency on the alphabet size, with randomized $\cO(n \log n)$ and $\cO(n \log m)$ solutions being proposed in~\cite{Indyk98fasteralgorithms} and~\cite{Kalai02efficientpattern-matching} respectively. Deterministic $\cO(n \log m)$ solutions were proposed soon after, first by Cole and Hariharan~\cite{Cole02verifyingcandidate} and then simplified by Clifford and Clifford~\cite{Clifford:2007:SDW:1222235.1222507}. The only known lower bound for the worst case of this problem is due to Muthukrishnan and Palem~\cite{Muthukrishnan:1994:NSA:195058.195457} who showed that in the worst case the problem is equivalent to computing boolean convolutions.

The indexing version of the problem was studied first in~\cite{Iliopoulos_patternmatching} where an index supporting queries in $\cO(m + \alpha)$ time was presented where wildcards are allowed in the pattern. A short coming of this approach is that in the worst-case $\alpha$ may be $\Theta(hn)$ where $h$ is the number of groups of wildcards. In~\cite{Cole:2004:DMI:1007352.1007374} Cole \textit{et al.} presented an index which given a text with $k$ wildcards and an integer $d$, allows searching for any pattern with at most $d$ wildcards. For a pattern containing $g \leq d$ wildcards, the matching takes $\cO(m + 2^g \log^k n \log\log n + occ)$ 
time\footnote{We use the notation of~\cite{Lam} as it is more understandable than ~\cite{Cole:2004:DMI:1007352.1007374}.}; when wildcards are restricted to either the pattern or the text the query time becomes $\cO(m + 2^g \log \log n + occ)$ and $\cO(m + \log^k n \log \log n + occ)$ respectively. A drawback of the index of Cole \textit{et al.} is that once the index has been built it can only be used to search for patterns with at most $d$ wildcards. 
Very recently a number of indexes were presented by Bille \textit{et al.}~\cite{DBLP:journals/corr/abs-1110-5236}, where a linear space index with query time $\cO(m + \sigma^g \log \log n + occ)$ and a linear query time index with space complexity $\cO(\sigma^{d^2}n \log^d \log n)$. The results of Billie \textit{et al.} can be improved using recent work on weighted ancestor queries~\cite{weightsuffixtrees}. In the area of linear size indexes for strings with wildcards, Lam \textit{et al.}~\cite{Lam} have presented indexes for a number of problems shown in Table~\ref{tab:lam}, where for a pattern $x$ consisting of strings $x_0, x_1,\dd , x_j$ interleaved by $j$ wildcards, $t_i$ is analogously defined for a text $t$ of length $n$, $occ(u,v)$ denotes the number of occurrences of $u$ in $v$, $\gamma = \Sigma_{j=1}^{\ell + 1} occ(t_j,x)$, $h$ is the number groups of consecutive wildcards in the text, $g$ is the total number of wildcards in the text, $\beta = \min_{1\leq i \leq h+1} \{occ(x_i,t)\}$.

\begin{table}
	\begin{tabular}{|l|c|c|r|}
	\hline \textbf{Problem} & \textbf{Query Time}  \\
	\hline  Wildcards in $t$ & $\cO(m \log n + \gamma + occ)$   \\ \hline
	Wildcards in $x$ & $\cO(m + h \beta )$   \\ \hline
	Wildcards in $t$ and $x$ & $\cO(m \log n + h\beta + \gamma + occ)$  \\ \hline
	Opt wildcards in $t$ & $\cO(m^2\log n + m \log^2 n + \gamma \log n + occ)$ \\ \hline
	Opt wildcards in $x$ & $\cO(m + gh\beta)$ \\ \hline
	Opt wildcards in $t$ and $x$ & $\cO(m^2 \log n + m \log^2 n + gh\beta + \gamma \log n + occ)$ \\ \hline
	\end{tabular}
	\caption{Properties of the indexes presented in \cite{Lam}}
	\label{tab:lam}
\end{table}

Succinct indexes have been presented in~\cite{sucwild} with a space usage of $((2 + o(1))n \log \sigma + \cO(n) + \cO(h \log n) + \cO(j \log j))$ bits for a text containing $h$ groups of $j$ wildcards in total.
The authors of~\cite{compresseda} proposed a compressed index where wildcards can only occur in the text with space usage $nH_y + o(n \log \sigma) + \cO(h \log n)$ bits, where $H_y$ is the $y$-th-order empirical entropy $(y = o(\log_\sigma n))$ of the text. The first non-trivial $o(n \log n)$ bit indexes were recently presented in~\cite{DBLP:journals/corr/LewensteinNV14}.

In this paper we focus on the average-case complexity of the problem. To the best of our knowledge no results are known.

\section{Preliminaries}

An \emph{alphabet} $\Sigma$ is a finite non-empty set, of size $\sigma$, whose elements are called \emph{letters}.
A \emph{string} on an alphabet $\Sigma$ is a finite, possibly empty, sequence of elements of $\Sigma$.
The zero-letter sequence is called the \emph{empty string}, and is denoted by $\varepsilon$. 
The \emph{length} of a string $x$ is defined as the length of the sequence associated with the string $x$, and is denoted by $|x|$. All strings of length $q$ are denoted by $\Sigma^q$ and refer to any $x \in \Sigma^q$ as a $q$-gram.
We denote by $x[i]$, for all $0 \leq i < |x|$, the letter at index $i$ of $x$.
Each index $i$, for all $0 \leq i <|x|$, is a position in $x$ when $x \neq \varepsilon$.
It follows that the $i$-th letter of $x$ is the letter at position $i-1$ in $x$, and that

\[
x = x[0\dd|x|-1].
\]

A string $x$ is a \emph{factor} of a string $y$ if there exist two strings $u$ and $v$, such that $y=uxv$.
Consider the strings $x,y,u$, and $v$, such that $y=uxv$. If $u=\varepsilon$, then $x$ is a \emph{prefix} of $y$.
If $v=\varepsilon$, then $x$ is a \emph{suffix} of $y$.

A \emph{wildcard} character is a special character that does not belong to alphabet $\Sigma$,
and matches with itself as well as with any character of $\Sigma$; it is denoted by $\phi$.
Two characters $a$ and $b$ of alphabet $\Sigma \cup \{\phi\}$ are said to \emph{correspond} (denoted by $a\approx^{\phi} b$) if they are equal or at least one of them is the wildcard letter.

Let $x$ be a non-empty string and $y$ be a string. We say that there exists an \emph{occurrence} of $x$ in $y$ or, more simply, that $x$
\emph{occurs in} $y$ when $x$ is a factor of $y$.
Every occurrence of $x$ can be characterised by a position in $y$. Thus we say that $x$ occurs at the \emph{starting position} $i$ in $y$ when $y[i\dd i+|x|-1]=x$. It is sometimes more suitable to consider the \emph{ending position} $i+|x|-1$. To be consistent with previous works on pattern matching with wildcards we consider the word RAM model of computation with word size $\Omega(\log n)$
In this paper the problems we consider are the following.

\begin{problem}[Wildcards in the Text]\label{prob:wt}
Given a text $t$ of length $n$ drawn from $\Sigma \cup \{ \phi \}$, and a pattern $x$ of length $m$ drawn from $\Sigma$. Find all $i$ such that $t[i \dd i+m-1] \approx^{\phi} x[0 \dd m-1]$.
\end{problem}

\begin{problem}[Wildcards in the Pattern]\label{prob:wp}
Given a text $t$ of length $n$ drawn from $\Sigma$, and a pattern $x$ of length $m$ drawn from $\Sigma \cup \{ \phi \}$. Find all $i$ such that $t[i \dd i+m-1] \approx^{\phi} x[0 \dd m-1]$.
\end{problem}

\section{Background on Average-case Analysis}
Here we give some background information on the literature concerning average-case complexity. The term \textit{average-case} has been used to refer to various different assumptions when discussing online pattern matching in strings, here we discuss these different assumptions and justify the model we use.

In the literature sometimes it is assumed that the pattern is randomly drawn from the alphabet, whilst others consider that the pattern is arbitrary.
An arbitrary pattern is assumed in~\cite{Lecroq94,Yao,Knuthf74fastpattern},
however, in later work such as~\cite{gonz,Fredriksson05flexiblemusic,Baeza-Yates:1989:ASS:74697.74700,Fredriksson:2004:ASM:1005813.1041513,Baeza-Yates98newand,Navarro:2002:FPM:571024,approxtriesearch} the assumption of a random pattern is made.
Clearly the notion of average-case complexity with arbitrary patterns is stronger than with random patterns and that is what we consider here.

Something else we consider is the fixed or adaptive nature of the sequence of probing positions applied to the text\footnote{The order characters are read in the text.}. We refer to the sequence of probing positions as the \textit{inspection scheme}. In~\cite{Yao} it was shown that having a predetermined inspection scheme negatively effects the runtime of exact pattern matching algorithms for $m<n<2m$ when compared with an adaptive inspection scheme. 
In this paper we explore the effect this property can have on the average-case performance of algorithms for pattern matching with wildcards and refer to algorithms which examine the character inside a window in a fixed order as \textit{fixed algorithms}. For the purpose of showing lower bounds in this paper we consider a simplified model of computation for fixed algorithms and define them as follows.

\begin{definition}[Fixed Algorithm]
Consider $t$ partitioned into non-overlapping blocks of size $2m$.
Let $(i_1, i_2, \dd, i_{2m})$ be an arbitrary but fixed permutation of 
$(0, 1, 2,3, \dd,2m-1)$. An algorithm is \textit{fixed} with respect to $(i_1, i_2, \dd, i_{2m})$ if, for every block, the sequence of probing positions is $(i_1, i_2, \dd, i_{2m})$. We wish to find all factors of $t$ corresponding to $x$ that are entirely within a block.
\end{definition} 

\noindent We consider the following simplified definition of \textit{non-fixed} algorithms.

\begin{definition}[Non-fixed Algorithm]
Consider $t$ partitioned into non-overlapping blocks of size $2m$.
An algorithm is \textit{non-fixed} if characters of $t$ can be inspected in any order. We wish to find all factors of $t$ corresponding to $x$ that are entirely within a block.
\end{definition}

We show a tight upper bound on the best performance for \textit{fixed} algorithms and for \textit{non-fixed} algorithms we show upper and lower bounds which match within a logarithmic factor for all but the most extreme values of $g$. The upper bounds for fixed and non-fixed algorithms are quite different.
 
It is important to point out that when discussing the average-case complexity of online string matching problems it is customary to make a distinction between time taking to preprocess the pattern and the \textit{search time}. \textit{average-case optimal} customarily refers to achieving the optimal search time, not necessarily considering the preprocessing time required to achieve it.\\

\noindent {\bf Our Contribution:} In this article, we present fast average-case algorithms for pattern matching with wildcards restricted to the text or pattern. We present algorithms with low preprocessing and average-case search time $\cO(\frac{n\log_\sigma m}{m})$ and $\cO(\frac{n(g+\log_\sigma m)}{m-g})$ for Problems~\ref{prob:wt} and~\ref{prob:wp} which are optimal in the non-fixed and fixed models respectively. We show an algorithm with optimal average-case search time for Problem~\ref{prob:wp} in the non-fixed model that has complexity between $\cO(\frac{n\log_{\sigma} m}{m-g})$ and $\cO(\frac{n\log_{\sigma} m \log_2 m}{m})$.\\

\section{Algorithms}\label{sec:wildalg}
In the following section we present a number of filtering algorithms for pattern matching in the presence of wildcards. The presented algorithms consist of two distinct
schemes: the \textit{filtering} scheme, which determines if the currently considered text
window \textit{potentially} has a valid occurrence; in case the window \textit{may} contain a
valid occurrence, we are required to check the window for valid occurrences of
the pattern; this is done through the \textit{verification} scheme.

Intuitively, the algorithms consider a {\em sliding window} of length $2m$ of the text, and reads $q$-grams backwards from the centre of the window until it is likely to have found a difference in every possible occurrence, allowing us to skip the window. That is, we wish to make the probability of a verification being triggered sufficiently unlikely whilst also ensuring we can shift the window a reasonable amount.

The verification scheme used in this algorithm consists of naively checking all possible alignments of the pattern against the text. Clearly each check takes no more than $\cO(m)$ time and there are $m$ possible start positions for a window of size $2m$ so $\cO(m^2)$ in total. For the rest of the article we refer to this verification scheme as $\textsf{VER}(i,x)$ where $i$ is the start of the window and $x$ is the pattern.

For the rest of the article we assume that the text $t$ is of length $n$ and is random and uniformly drawn from $\Sigma$ or $\Sigma \cup \{ \phi \}$, depending on the problem, and that $\Sigma$ is a finite but not necessarily constant alphabet.

\subsection{Filtering Scheme}
The filtering scheme of the presented algorithm requires the preprocessing and
indexing of the pattern $x$. We first present the preprocessing required and then
present the searching technique itself.\\

\noindent \textbf{Preprocessing.} We outline two indexing schemes that will be used in this paper. The first is a basic $q$-gram index adapted to allow wildcards in the query, which simply considers if a matching $q$-gram exists in the indexed text. To build this index we generate all strings of length $q$ from the alphabet $\Sigma \cup \{\phi \}$, and check each against the set of $q$ length factors from the pattern. Since there are $(\sigma + 1)^q$ different generated strings, at most $m$ factors of length $q$ in the pattern and each check requires $\cO(q)$ time, the total time spent is $\cO(m q (\sigma +1)^q)$. We can store the results of this processing in a binary array where each string is converted to a numerical representation for efficient lookup. A dictionary built in this way has a search time of $\cO(q)$. We refer to this scheme as $\textsf{q-Basic}$. 

The second indexing scheme is based on the simple $q$-gram index but allows wildcards in the text and for multiple queries allows us to enforce a weak ordering on the $q$-grams. The intuition for this scheme is that if we index a text with $g$ wildcards, it may be the case that all of these wildcards occur sequentially. In this situation if $q<g$ then the basic $q$-gram index will report a match for every single query. This makes it impossible for us to use the basic $q$-gram index as part of a filtering scheme unless $q>g$ which gives an exponential dependency on $g$. To get around this issue we build a $q$-gram index by considering every length $q$ factor of the pattern and generating all matching strings over $\Sigma$. Let $\textsf{M}[0\dd \sigma^q -1,0 \dd m-1]\leftarrow 1$ be a 2-d array and $q_1$ be the $q$-gram starting at $x[i]$, then for every $s \in \Sigma^q$ such that $s=q_1$ we store $\textsf{M}[\rho(s)][i] \leftarrow 0$ where $\rho(s)$ is the numerical representation of $s$.. This procedure is done for every $q$-gram in the pattern. Clearly for each $q$-gram considered there are at most $\sigma^q$ matching factors and no more than $m$ $q$-grams will be considered giving $\cO(m\sigma^q)$ time. Similarly \textsf{M} is of size at most $\sigma^q$ and each list is not longer than $m$ giving $\cO(m\sigma^q)$ space. We preprocess each array in $\textsf{M}$ for next smaller value 	queries. Next smaller value queries take $\cO(1)$ time after linear preprocessing~\cite{Barbay201226}. We refer to this scheme as $\textsf{q-Weak}$.

Consider a sequence $s_1,s_2, \dd, s_{\ell}$ of $\ell$ consecutive, non-overlapping $q$-grams read from a text $t$, we make use of the observation that any occurrence of the pattern containing those $q$-grams must contain those $q$-grams in the same order that they appear in the text. Whilst it is possible to check if all possible start locations have the $q$-grams occurring consecutively this search may be costly. We weaken the restriction on $q$-grams and require only that for some $s_i$ occurring at position $j$ in the pattern that there exists an occurrence of $s_{i+1}$ at some $h > j+q$. If there are multiple occurrences then we take the minimum starting position greater than $j+q$. Given a $\textsf{q-Weak}$ index, $\ell$ query $q$-grams such that $s_{i}$ occurs at position $j$, this corresponds to checking if $s_{i+1}$ exists and then performing a next smaller value query on $\textsf{M}[\rho(s_{i+1})][j+q]$. We denote this query by $\textsf{Weak-Order-Query}(s_1,s_2, \dd, s_{\ell})$ and with this we get the following.

\begin{lemma}
Let $x$ be a pattern of length $m$ with $g$ wildcards, $t$ be a text of length $n$ with no wildcards and let $s_1,s_2, \dd, s_{\ell}$ be a sequence of $\ell$ consecutive, non-overlapping $q$-grams read from $t$. Given a \textsf{q-Weak-Fast} index of $x$ we can perform $\textsf{Weak-Order-Query}(s_1,s_2, \dd, s_{\ell})$ in $\cO(\ell q + \ell)$.
\end{lemma}

\subsection{Wildcards in Text Only}
We begin by considering the problem where wildcards may appear only in the text and refer to the algorithm of this section as Algorithm \textsf{Wt}. 
First we will establish the probability of a random string of length $x \log_{\frac{\sigma + 1}{2}} m$ containing wildcards matching some factor of the pattern. This choice of length will become clear later.

\begin{lemma} \label{rand}
Let $u$ be a random string of length $x \log_{\frac{\sigma + 1}{2}} m$ over $\Sigma \cup \{ \phi\}$ and let $v$ be a string of length $m$ over $\Sigma$. The probability of $u$ matching a factor of $v$ is at most $\frac{1}{m^{x-1}}$.
\end{lemma}
 
Applying index $\textsf{q-Basic}$ we can see that for $q=3 \log_{\frac{\sigma + 1}{2}} m$ this is a polynomial preprocessing scheme and the maximum exponent occurs when $\sigma = 2$. For $\sigma = 2$ we have that $3^{3 \log_{1.5}m} \approx m^{8.13}$ and so the total preprocessing is $\cO(m^{9.13}\log m)$. Asymptotically on the alphabet size this becomes $\cO(m^{3}\log m)$.

Consider a sliding window of length $m$ placed over the text and that for each window we will check if the suffix of length $3 \log_{\frac{\sigma + 1}{2}} m$ corresponds with any factor of the pattern. If the suffix corresponds with a factor of the pattern there may be a match and we are required to verify this window. To verify the window we run algorithm $\textsf{VER}(i,x)$ in time $\cO(m^2)$. For each window on the text we have a total possible running time of $\cO(m^2 + \log_{\frac{\sigma +1}{2}} m)$. The probability of the suffix corresponding with a factor of the pattern is $1/m^2$ by Lemma~\ref{rand}, so the expected time spent at each window is $\cO(\log_{\frac{\sigma +1}{2}} m)$ or $\cO(\log_\sigma m)$. If the window is verified then it is possible to shift by $m$ characters, otherwise the window is not verified and can we shift by $m-q$ characters. This means there there are at most $\frac{n}{m-q}$ windows, each has an expected runtime of $\cO(\log_\sigma m)$, giving us an overall average-case running time of $\cO(n\log_\sigma m /m)$. From the above discussion, and noting that Yao's lower bound~\cite{Yao} for exact string matching is $\cO(n \log_\sigma m /m)$, we get the following result:

\begin{theorem}\label{thrm:wt}
Algorithm $\textsf{Wt}$ has optimal average-case search time $\cO(n \log_\sigma m /m)$ with $\cO((\sigma +1)^{3 \log_{\frac{\sigma + 1}{2}}m}m\log_{\sigma} m)$ preprocessing time and $\cO((\sigma +1)^{3 \log_{\frac{\sigma + 1}{2}}m})$ space.
\end{theorem}


\subsection{Wildcards in the Pattern}

In this section we consider the case where wildcards may appear in both the pattern and text. For a pattern $x$ we denote the number of wildcards in the pattern by $g$. We refer to the algorithm of this section as Algorithm \textsf{Wp}.

We could use dictionary $\textsf{q-Basic}$ but for this problem we would have that $q > g$. This means that the space complexity of the dictionary would become $\Omega((\sigma+1)^{g})$ which is far too high. Instead we will use dictionary $\textsf{q-Weak-Fast}$. We now establish the probability of random string matching a pattern with $g$ wildcards.

\begin{lemma}\label{lem:wp}
Let $u$ be a random string of size $g + x \log_{\sigma} m$ over $\Sigma$ and let $v$ be a string of size $m> g+ \log_{\sigma} m$ over $\Sigma \cup \{ \phi \}$ with $g$ wildcards.
The probability of $u$ matching a factor $v$ is no more than $\frac{1}{m^{x-1}}$.
\end{lemma}

For this algorithm we create a sliding window on the text of length $2m$ and build a $\textsf{q-Weak-Fast}$ index over $x$ with $q=3 \log_\sigma m$. For each window on the text we denote the start position by $i$ and read $\frac{g}{q} + 1$ $q$-grams starting at $t[i+m-g-q-1]$ and perform a $\textsf{Weak-Order-Query}$ on them.

If the $\textsf{Weak-Order-Query}$ returns true then we perform $\textsf{VER}(i,x)$ to verify all possible start positions in the current window and shift by $m$ positions. If the $\textsf{Weak-Order-Query}$ returns false we shift by $m-g-q$. The minimum shift the algorithm makes is $m - g - 3\log_{\sigma} m$, so there will be at most $\frac{n}{m-g-3 \log_{\sigma} m}$ windows on the text and at each window we may do $\cO(m^2+ g + \log_{\sigma} m)$ work in the worst case. 
The probability that we will need to verify a window is $1/m^2$ by Lemma~\ref{lem:wp}, this gives us an expected time of $\cO(g + \log_{\sigma} m)$ per window.  

For the algorithm to achieve the claimed runtime it must be the case that $\frac{n}{m-g-3 \log_{\sigma} m} = \cO(\frac{n}{m})$. To satisfy this it follows that $g+3\log_{\sigma} m \leq cm$ for some $0<c <1$. This places the following condition on the wildcard ratio of our algorithm:

$$\frac{g}{m} < c - \frac{3 \log_{\sigma} m}{m}$$

\noindent We also have an additional restriction, we must be able to guarantee that we are reading enough new random characters after each shift that Lemma~\ref{lem:wp} still holds. This places the additional restriction that $m$ must be at least twice the length of the shortest shift. So it must hold that $m > 2(g + 3\log_{\sigma} m)$ to ensure that in all cases we read enough new $q$-grams from a window for the above analysis to hold. After rearrangement this places the following restriction on our algorithm:

$$\frac{g}{m} < \frac{1}{2} - \frac{3 \log_{\sigma} m}{m}$$

\noindent Clearly the second condition places the strictest condition on the wildcard ratio. From the above discussion we achieve the following result:

\begin{theorem}\label{thrm:wpr}
Algorithm $\textsf{Wp}$ has average-case search time $\cO(\frac{n(g+ \log_{\sigma} m)}{m})$ with $\cO(m^4)$ preprocessing time and $\cO(m^4)$ space, for $\frac{g}{m} < \frac{1}{2} - \frac{3 \log_{\sigma} m}{m}$.
\end{theorem}

The wildcard ratio we specify is quite permissive as $\frac{3 \log_{\sigma} m}{m}=o(1)$ so for any $g/m < 1/2$ it is possible to pick a sufficiently large value of $m$ such that the algorithm can run in the claimed running time.
In the following theorem we show that for any fixed algorithm it is impossible to do any better than this. We show that for any integer $g$, there exists a lower bound of $\Omega(\frac{ng}{m})$ character inspections for any fixed algorithms.

\begin{theorem}\label{thrm:wp}
Algorithm $\textsf{Wp}$ has average-case search time $\cO(\frac{n(g+ \log_{\sigma} m)}{m})$ and no fixed algorithm can do better.
\end{theorem}


\section{A General Lower Bound}\label{sec:low}
In the previous section we have considered the average-case complexity of each algorithm. Fixed algorithms consider the characters in each window of the text in a fixed order, an approach that is ubiquitous in string algorithms. We have shown that for any fixed inspection scheme there exists patterns that perform badly on average.
In this section we consider non-fixed algorithms and derive an average-case lower bound for any algorithm solving problem~\ref{prob:wp} with an arbitrary pattern and any value for $g$. This bound gives a provable separation in complexity between exact string matching and wildcard matching. Clearly a lower bound for this problem also lower bounds the problem where wildcards appear in both the pattern and the text. 

When considering this problem the order characters are inspected becomes very important.
Some inspection schemes do not have much effect on the expected number of candidates not yet 
ruled out. It is the case that inspections that, should wildcards not be present, would lead to a large reduction in the expected number of candidates may give very little information when wildcards exist.

Consider that we have a pattern of length $m$ with $g<m$ wildcard characters and a text of length $n$. Partition the text into non-overlapping \textit{blocks} of size $2m$, and only consider that we must report all matches within each block. This is an optimistic assumption as this excludes those matches which overlap two blocks. In the following section we will determine a lower bound for the number of character inspections required for \textit{one} block. The lower bound for the general problem can then be derived.

For each block we call $0, \dd, m-1$ possible starting positions \textit{candidates} and when we inspect a character from the block we call this a \textit{block access}. The candidates affected by a block access are \textit{intersected} by it. Given a block access to position $z$ in block $b$ we can only rule out candidate $c$ if there exists some $y$ such that $x[y] \neq b[z]$ and $c+y = z$. 
For all non-wildcard positions intersected by a given block access $i_j$, there is a probability of at most $1/\sigma$ that the candidate will not be ruled out. For those candidates where this block access intersects a wildcard there is probability $1$ it will not be ruled out. We outline a few optimistic assumptions used in our analysis. 

\begin{itemize}
	\item Any block access intersects all $m$ candidates. 
	\item Intersections are distributed uniformly across all candidates.
\end{itemize}

\noindent The effect of this is that $m-g$ candidates have a chance of being ruled out at every block access. After $k$ block accesses in this model we have made $(m-g)k$ intersections and 
we assume that these are distributed uniformly across all $m$ candidates. This is optimistic as
this means that we always intersect candidates with the highest probability of occurrence, something which may not actually be possible. This means that we may only overestimate the number of candidates ruled out and our result is a lower bound. Clearly the expected number of candidate not ruled out is then be evaluated as follows:

$$\displaystyle\sum_{i=0}^{m-1} \frac{1}{\sigma^{\frac{(m-g)k}{m}}} = \frac{m}{\sigma^{\frac{(m-g)k}{m}}}$$

\noindent For each candidate we need to either rule it out as a possible starting position or declare a match. So the optimal is to determine when we would expect to have ruled out every candidate position or read $2m$ characters. We minimise the following so that we expect to have at most one candidate left or until we have read all $2m$ positions. 

$$\frac{m}{\sigma^{\frac{(m-g)k}{m}} }\leq 1$$

\noindent Rearranging this we get the following:

$$\log_\sigma m\leq \frac{(m-g)k}{m}$$

$$\frac{m\log_\sigma m}{m-g}\leq k$$

\noindent Now we know that the lower bound for each block is $\Omega(\frac{m \log_\sigma m}{m-g})$ and there are $n/2m$ blocks. The result below then follows:

\begin{theorem}\label{thrm:lowb}
The average-case lower bound for wildcard matching with wildcards only in the pattern is $\Omega(\frac{n \log_\sigma m}{m-g})$. 
\end{theorem}

\noindent Clearly this results suggests that pattern matching with wildcards is on average asymptotically harder than exact string matching when $g = m - f(m)$, where $f(m)$ is some function asymptotically smaller than $\cO(m)$.




Now we discuss a strategy for inspecting positions of a block and show that a greedy scheme performs an optimal number of character comparisons. By greedy we mean that at each step the block access which would most greatly reduce the expected number of remaining positions is chosen. For a candidate $i$ let $\mathbb{a}_i$ be the number of times it has been intersected at a non-wildcard position. Now for each position in a block $0 \leq i < 2m$ let $\mathcal{B}_i$ be the set of candidates that the block access $i$ intersects.

The effect on the expected number of candidates not ruled out by inspecting some position $\ell$ is given by the following. Let $\mathcal{U}= \{0,1, \dd, m-1\} \textbackslash \mathcal{B}_\ell$ and $\mathbb{a}_i$ denote the number of times candidate $i$ has been intersected before the block access to $\ell$.

$$\displaystyle\sum_{i=0}^{m-1} \frac{1}{\sigma^{\mathbb{a}_i}} - \displaystyle\sum_{j\in \mathcal{U}} \frac{1}{\sigma^{\mathbb{a}_j}} - \displaystyle\sum_{h\in \mathcal{B}_\ell} \frac{1}{\sigma^{\mathbb{a}_h + 1}}$$ 

\noindent The greedy inspection maximises the last two terms of the above summation at each step. The intuition behind this scheme is based on our proof of the lower bound presented above. In the proof we saw that evenly distributing accesses across all candidates minimises the expectation. The greedy scheme attempts to simulate this behaviour by picking the access which most minimises the probability at each step.
We now show that this is in fact optimal.




\begin{theorem}\label{thrm:inspec}
The greedy inspection scheme performs an optimal number of character comparisons.
\end{theorem}

We have shown that a greedy inspection scheme minimises the number of character inspections required for any $g$, but this does not give us any explicit upper bound. We now give an upper bound on the average-case search time significantly better than that in algorithm $\textsf{Wp}$ for all but the most extreme values of $g$.

\section{An Improved Upper Bound}
As before consider that we have a pattern of length $m$ which contains $g < m$ wildcard characters
and a text of length $n$. We partition the text into non-overlapping blocks of size $2m$	 and only consider that we have to report all matches
within each block. Although this is an optimistic assumption, we will show how to convert this to a true upper bound later in this section. In the rest of this section we will determine an upper bound for the number of character inspections required for one block. First we recall the definition of the $\varepsilon$-dense set cover problem along with a known approximation results.

\begin{problem}
Given $\varepsilon>0$, a set of elements $\textsf{U} = \{ 1,2,3, \dd,r \}$, a family \textsf{S} of $\ell$ sets such that every element of $\textsf{U}$ occurs in $\varepsilon r$ sets and the union of \textsf{S} equals $\textsf{U}$. Find the minimum number of sets from \textsf{S} such that their union is \textsf{U}.
\end{problem}

\noindent The general set cover problem is known to be \textsf{NP-hard}, however the following approximation result is known for the $\varepsilon$-dense set cover problem.

\begin{lemma}[\cite{Karpinski:1996:ADC:895399}]
There exists an approximation algorithm for the $\varepsilon$-dense set cover problem with output size $\log_{\frac{1}{1-\varepsilon}} r$ where $\vert \textsf{U} \vert = r$.
\end{lemma}

\noindent We define the following family of sets for $1\leq j < 2m$ $$S_j = \{\text{ }i\in \{1, 2, \dd, m\}\text{ }:\exists\text{ }i<j\text{ and }c>0 \text{ s.t. } x[c] \neq \phi\text{ and }c + i =j  \}$$

Our goal in this section is to determine how many inspections it takes to make the probability of any candidate not being rules out smaller enough to make verification time negligible. One way to do this is to ensure that all candidates have been intersected at least $3\log_{\sigma} m$, causing the probability one is not ruled out to be at most $1/m^2$. The difficulty in deriving the upper bound is we must explicitly consider how each candidate is affected by a block access. To do this we use the above result on $\varepsilon$-dense set covers. By the definition of $\textsf{S} = S_1,\dd, S_{2m}$ it is clear that each element occurs in $m-g$ sets, so the problem can be seen as an $\frac{m-g}{2m}$-dense set cover problem. We will apply the $\varepsilon$-dense set cover problem, remove the approximate set cover and apply the $\varepsilon$-dense set cover problem again until we achieve $3\log_{\sigma} m$ intersections per candidate. We now define a family of functions whose argument is the pattern $x$ which relate to repeated applications of the $\varepsilon$-dense set cover problem applied on \textsf{S}.	

$$\mathcal{F}^0(x) = \log_2 \frac{2m}{m + g}$$

$$\mathcal{F}^i(x) = \log_2 \frac{2m}{m + g + \frac{\log_2 m}{\mathcal{F}^0(x)} + \frac{\log_2 m}{\mathcal{F}^1(x)} + \dd + \frac{\log_2 m}{\mathcal{F}^{i-1}(x)}}$$

\noindent We are now in a position to define a density function which gives the density of the remaining sets after $i$ applications of the $\varepsilon$-dense set cover problem.

$$\mathcal{D}^0(x) = \frac{m -g}{2m}$$

$$\mathcal{D}^i(x) = \frac{m -g - \frac{\log_2 m}{\mathcal{F}^0(x)} - \frac{\log_2 m}{\mathcal{F}^1(x)} - \dd - \frac{\log_2 m}{\mathcal{F}^{i-1}(x)}}{2m}$$

\noindent The number of characters inspected to guarantee $i>0$ intersections per candidate is then given by

$$\mathcal{G}^i(x) = \frac{\log_2 m}{\mathcal{F}^0(x)} + \frac{\log_2 m}{\mathcal{F}^2(x)} +  \dd + \frac{\log_2 m}{\mathcal{F}^{i-1}(x)}$$

\noindent We now show a number of properties of the above functions. We wish to show that for some integer $i\geq 0$, a constant $0<\epsilon<1$ and $g + \frac{\log_2 m}{\mathcal{F}^0(x)} + \frac{\log_2 m}{\mathcal{F}^1(x)} + \dd + \frac{\log_2 m}{\mathcal{F}^{i-1}(x)} <\epsilon m$.

$$ \log_2 \frac{2}{1 +\epsilon} \leq \lim_{m \to \infty} \mathcal{F}^i(x)  \leq \log_2 2$$

\noindent We proceed by induction on $i$. Let $i=0$ and $g<\epsilon m$ then clearly 

$$ \log_2 \frac{2}{1 +\epsilon} \leq \lim_{m \to \infty} \frac{2m}{m +g} \leq \log_2 2$$

\noindent Let $i=k$, $0<\epsilon <1$ and assume for all natural numbers less than $k$ that the induction hypothesis holds. We focus on the lower bound as the upper bound is obvious. Let $i=k+1$ and $g + \frac{\log_2 m}{\mathcal{F}^0(x)} + \frac{\log_2 m}{\mathcal{F}^1(x)} + \dd + \frac{\log_2 m}{\mathcal{F}^{k}(x)} < \epsilon m $. Applying the induction hypothesis this is at least $g + \underbrace{\frac{\log_2 m}{\log_2 \frac{2}{1 +\epsilon}} + \frac{\log_2 m}{\log_2 \frac{2}{1 +\epsilon}} + \dd + \frac{\log_2 m}{\log_2 \frac{2}{1 +\epsilon}}}_{k+1 \text{ times}} < \epsilon m $ as $m \rightarrow \infty$ and therefore.
$$\lim_{m \to \infty} m+ g + \frac{\log_2 m}{\mathcal{F}^0(x)} + \frac{\log_2 m}{\mathcal{F}^1(x)} + \dd + \frac{\log_2 m}{\mathcal{F}^{k}(x)} \leq \epsilon m$$

\noindent It then follows that 
$$\log_2 \frac{2}{1 +\epsilon} \leq \lim_{m \to \infty} \mathcal{F}^{k+1}(x) \leq \log_2 2$$

\noindent With this result we get that for sufficiently large $m$ and $g + \frac{\log_2 m}{\mathcal{F}^0(x)} + \frac{\log_2 m}{\mathcal{F}^1(x)} + \dd + \frac{\log_2 m}{\mathcal{F}^{i-1}(x)} < \epsilon m$.

$$\mathcal{G}^i(x) = \cO(i \log_2 m)$$

\noindent Recall that we wish to intersect each candidate at least $3\log_{\sigma} m$ so for $i = 3\log_{\sigma} m +1$ we get the number of inspected characters is $\mathcal{G}^{3 \log_{\sigma} m +1}(x) = \cO(\log_{\sigma}m \log_2 m)$ for sufficiently large $m$. This upper bound is for a block of the text of size $2m$ and assumes all character inspections can be considered as independent tests. To convert this to a general bound for a text of size $n$ consider the text partitioned into blocks of size $2m$ that overlap by $m$ characters. After inspecting $\cO(\log_{\sigma}m \log_2 m)$ characters we expect that we can discard the entire block and can shift by $m$ characters. It may be the case that we have already read some of the characters we need to inspect, but as long as the window is twice the size of $\cO(\log_{\sigma}m \log_2 m)$ the analysis still holds. Computing the greedy inspection scheme can be trivially done in $\cO(m^3)$ and an index can be built by building a finite state machine memorising the candidates still valid for any possible character read in the order specified, this takes no more than $\sigma^{\cO(\log_{\sigma}m \log_2 m)}$ which is polynomial for $\sigma=\cO(1)$ and as $\sigma^{\cO(\log_{\sigma}m \log_2 m)}= 2^{\cO(\log_{\sigma} m(\log_2 m)^2)}$ is quasi-polynomial for larger alphabets, therefore.

\begin{theorem}\label{thrm:optopt}
For any constant $0 <\epsilon<1$ there exists an algorithm for Problem \ref{prob:wp} with average-case search time $\cO(\frac{n\log_{\sigma} m \log_2 m}{m})$ when $g + \frac{\log_2 m}{\mathcal{F}^0(x)} + \frac{\log_2 m}{\mathcal{F}^2(x)} + \dd + \frac{\log_2 m}{\mathcal{F}^{3\log_2 m }(x)} < \epsilon m$.
\end{theorem}

\noindent Combining all the above we know that there exists an algorithm that has the following bounds on the number of character required to be inspected. Let \textsf{OPT} denote the number of characters inspected by the optimal scheme with the specified conditions on $g$.

$$\cO\Big(\frac{n\log_{\sigma} m}{m-g}\Big) \leq \textsf{OPT} \leq \cO\Big(\frac{n\log_{\sigma} m \log_2 m}{m}\Big)$$

\section{Conclusions and Future Work}\label{sec:concu} 
In this paper we have investigated the average-case complexity of two wildcard matching problems.
We have considered two models of computation and have shown a big jump in the complexity of the two. The question of a tight bound on the search complexity of pattern matching with wildcards remains open. We have shown an algorithm which has optimal average-case search time and upper/lower bounds on the time complexity that are only different by a logarithmic factor. This lower bound shows the first provable separation in time complexity between wildcard matching and exact matching in the word RAM model. Although~\cite{Muthukrishnan:1994:NSA:195058.195457} showed that in the worst case the problem is equivalent to computing boolean convolutions and many believe this cannot be solved in linear time, no superlinear lower bound on this is known. We conjecture that the optimal number of character inspections is in fact the lower bound stated in Theorem~\ref{thrm:lowb}.




\bibliographystyle{plain}
\bibliography{pnfw.bib}

\section*{Appendix}

\paragraph*{Proof of Lemma 4}

\begin{proof}
The probability of a character from $\Sigma$ matching a randomly picked character of $\Sigma \cup \{ \phi \}$ is $\frac{2}{\sigma +1}$. The probability of $r$ characters matching in this way is given by $(\frac{2}{\sigma +1})^r$, by setting $r = x \log_{\frac{\sigma +1}{2}} m$ we get that $(\frac{2}{\sigma +1})^r = \frac{1}{m^{x}}$. There are at most $m - x\log_{\frac{\sigma+1}{2} }m +1$ factors of this length in $v$ and so the probability of occurrence is no more than $\frac{1}{m^{x-1}}$.
\end{proof}

\paragraph*{Proof of Lemma 6}

\begin{proof}
In the worst case all $g$ wildcards of $v$ appear in a single factor of size $g + x \log_{\sigma} m$, this factor contains $x \log_\sigma m$ positions which are not wildcards and the probability that these match the corresponding characters in $u$ is no more than $\frac{1}{m^{x}}$. For any factor with less than $g$ wildcards the probability is $ \leq \frac{1}{m^{x}}$. Pessimistically assume all factors of $v$ of length $g + x\log_{\sigma} m$ match with probability $\frac{1}{m^{x}}$, there are $m- (g + x\log_{\sigma} m)$ factors of this length so the probability is no more than $\frac{1}{m^{x-1}}$. 
\end{proof}

\paragraph*{Proof of Theorem 2}

\begin{proof}
Recall that the lower bound for exact string matching is $\Omega(n \log_\sigma m / m)$. We now show that any fixed algorithm has a lower bound of $\Omega(\frac{ng}{m})$. 

Assume the text is partitioned into non-overlapping blocks of size $2m$ and that we only want to find occurrences contained entirely within these blocks. Let $\pi^{2m}$ denote all the permutations of $(0,1, \dd , 2m-1)$ and assume that we examine the characters of each block in the same fixed but arbitrary order $(i_0, i_1, \dd, i_{2m-1}) \in \pi^{2m}$. We can construct patterns, for any $g < m$, such that we must examine at least $g+1$ characters before all start positions can be ruled out in the following way. If for $0 \leq j < g$ all $i_j$ occur within a range of $m$ positions then we place the wildcards in positions $i_0, i_1, \dd, i_{g-1}$ of the pattern. Otherwise for $0 \leq j < g$ and $i_j<m$ place a wildcard at position $i_j$. Any remaining wildcards may be placed anywhere in the pattern; the remaining positions of the pattern are random characters from $\Sigma$. 
After inspecting characters $i_0, i_1, \dd, i_{g-1}$ of the block at least the first position can neither be ruled out nor declared as a match.
Combining the lower bound of Yao and this we see that any fixed algorithm has a lower bound of $\Omega(n(g + \log_\sigma m)/m)$ for this problem. Algorithm $\textsf{Wp}$ runs in average-case time $\cO(n (g + \log_\sigma m) /m)$, matching the lower bound; therefore the algorithm is optimal in the family of fixed algorithms. 
\end{proof}

\paragraph*{Proof of Theorem 5}

\begin{proof}
The optimal number of character comparisons is achieved by a scheme minimising the number of block accesses needed to expect that less than 1 candidate remains. An inspection scheme $\mathbb{L}_k$ is simply a $k$ subset of the probing positions $ \{0,1,2,\dd,2m-1 \}$. For some inspection scheme $\mathbb{L}_k$ we denote by $\textsf{E}[\mathbb{L}_k]$ the expected number of remaining candidates for inspection scheme $\mathbb{L}_k$.
We proceed by induction on the number of block access and claim that the greedy scheme is an optimal scheme. Let $\mathbb{G}_{i} = \{ \mathbb{g}_1,\mathbb{g}_2, \dd, \mathbb{g}_{i} \}$ be the first $i$ block accesses made by the greedy inspection scheme. $\mathbb{G}_{i+1}$ is then defined as $\mathbb{G}_{i} \cup \mathbb{g}_{i+1}$ where $\mathbb{g}_{i+1}$ is the block access with the biggest effect on the expected number of candidates not ruled out.
The base case is simple, we pick the block access which minimises the expected number of candidates not ruled out, by definition this is the minimum.


Assume that for some $i$ it is true that for the greedy scheme the number of expected candidates is the smallest possible after $i$ accesses. 
Let $\mathbb{K}_{i} = \{ \mathbb{k}_{1}, \mathbb{k}_{2}, \dd, \mathbb{k}_{i} \}$ be an arbitrary inspection scheme with $i$ block accesses. If $\mathbb{G}_{i} = \mathbb{K}_{i}$ then $\textsf{E}[\mathbb{G}_{i+1}] \leq \textsf{E}[\mathbb{K}_{i+1}]$ and we are done, otherwise we have a few cases to consider.

Consider $\mathbb{G}_{i+1}$ and $\mathbb{K}_{i+1}$, if $\mathbb{k}_{i+1} = \mathbb{g}_{i+1}=\ell$.
Let $\mathcal{B}_\ell$ be the set of candidates that the block access $\ell$ intersects. If the following holds:

$$\displaystyle\sum_{h\in \mathcal{B}_\ell} \frac{1}{\sigma^{\mathbb{k}_h }} \leq \displaystyle\sum_{h\in \mathcal{B}_\ell} \frac{1}{\sigma^{\mathbb{g}_h }}$$

\noindent Then so does the following:

$$\displaystyle\sum_{h\in \mathcal{B}_\ell} \frac{1}{\sigma^{\mathbb{k}_h + 1}} \leq \displaystyle\sum_{h\in \mathcal{B}_\ell} \frac{1}{\sigma^{\mathbb{g}_h + 1}}$$

\noindent Additionally, if the following is true:

$$\displaystyle\sum_{h\in \mathcal{B}_\ell} \frac{1}{\sigma^{\mathbb{k}_h }} \geq \displaystyle\sum_{h\in \mathcal{B}_\ell} \frac{1}{\sigma^{\mathbb{g}_h }}$$

\noindent Then so is the following:

$$\displaystyle\sum_{h\in \mathcal{B}_\ell} \frac{1}{\sigma^{\mathbb{k}_h + 1}} \geq \displaystyle\sum_{h\in \mathcal{B}_\ell} \frac{1}{\sigma^{\mathbb{g}_h + 1}}$$

\noindent Therefore, should $\mathbb{k}_{i+1} = \mathbb{g}_{i+1}$ then $\textsf{E}[\mathbb{G}_{i+1}]\leq \textsf{E}[\mathbb{K}_{i+1}]$.

Let $\mathbb{C} = \{\{ 0, 1, \dd 2m-1\}\textbackslash \mathbb{G}_i\} \textbackslash \mathbb{K}_i$ and $\mathbb{k}_{i+1} \neq \mathbb{g}_{i+1}$ if $\mathbb{g}_{i+1}, \mathbb{k}_{i+1} \in \mathbb{C}$ then $\textsf{E}[\mathbb{G}_{i+1}] \leq \textsf{E}[\mathbb{K}_{i+1}]$, otherwise either $\mathbb{g}_{i+1}\not\in \mathbb{C}$ or $\mathbb{k}_{i+1} \not\in \mathbb{C}$. 
If $\mathbb{k}_{i+1} \neq \mathbb{g}_{i+1}$ then either $\mathbb{k}_{i+1}\in \mathbb{G}_{i}$ or $\mathbb{g}_{i+1}\in \mathbb{K}_i$.	It is sufficient to show that there exists at least one possible choice for $\mathbb{g}_{i+1}$ that would cause $\textsf{E}[\mathbb{G}_{i+1}]\leq \textsf{E}[\mathbb{K}_{i+1}]$ in each case.\\

\noindent \underline{Case 1: $\mathbb{k}_{i+1}\in \mathbb{G}_i$}

\noindent Let $\mathbb{k}_u \in \mathbb{K}_i\textbackslash \mathbb{G}_i$, $\mathbb{k}_u$ must exist as $\mathbb{K}_i \neq \mathbb{G}_i$,
and consider the inspection scheme $\mathbb{K}^{\prime}_i = \{ \mathbb{K}_i\textbackslash \mathbb{k}_u\}\cup \mathbb{k}_{i+1}$. We claim that $\mathbb{g}_{i+1}=\mathbb{k}_u$ ensures optimality. By the induction hypothesis $\textsf{E}[\mathbb{K}^{\prime}_i] \geq \textsf{E}[\mathbb{G}_{i}]$. Now let  $\mathbb{g}_{i+1} = \mathbb{k}^{\prime}_{i+1} =\mathbb{k}_u$, by the above analysis for when $\mathbb{g}_{i+1} = \mathbb{k}^{\prime}_{i+1}$, $\textsf{E}[\mathbb{G}_{i+1}]\leq \textsf{E}[\mathbb{K}^{\prime}_{i+1}]$ and therefore $\textsf{E}[\mathbb{G}_{i+1}]\leq \textsf{E}[\mathbb{K}_{i+1}]$.\\
 
\noindent \underline{Case 2: $\mathbb{g}_{i+1}\in \mathbb{K}_i$}

\noindent If $\mathbb{k}_{i+1}\in \mathbb{C}$, the greedy scheme could also pick the block access $\mathbb{k}_{i+1}$ and remain optimal so it must be the case that $\textsf{E}[\mathbb{G}_{i+1}]\leq \textsf{E}[\mathbb{K}_{i+1}]$. If $\mathbb{k}_{i+1}\not\in \mathbb{C}$ then $\mathbb{k}_{i+1}\in \mathbb{G}$ and by case 1 the greedy scheme is optimal. Therefore $\textsf{E}[\mathbb{G}_{i+1}]\leq \textsf{E}[\mathbb{K}_{i+1}]$ and the statement is proved.
\end{proof}

\end{document}